\documentclass[showpacs,amsmath,onecolumn,showkeys,pra,11pt]{revtex4-1}
\usepackage{bm}         
\usepackage{ifpdf}
\usepackage{amssymb,lineno,amsfonts}
\usepackage{graphicx}   
\usepackage{bbm}
\usepackage{mathrsfs}
\usepackage{upgreek}
\usepackage{epstopdf}
\usepackage{setspace}
\usepackage{hyperref}
\usepackage[matrix,frame,arrow]{xypic}
\usepackage{float}
\usepackage{natbib}
\usepackage{color} 
\newcommand{\ket}[1]{\vert{#1}\rangle}

\newcommand{\outpr}[2]{\vert{#1}\rangle\langle{#2}\vert}

\newcommand{\expv}[1]{\langle{#1}\rangle}

\newcommand{\proj}[1]{\outpr{#1}{#1}}
\newcommand{\expec}[1]{\langle{#1}\rangle}

\newcommand{\tr}{\mathrm{Tr}}

\begin{document}
	
	\title{Quantum Correlations in NMR systems}
	\author{T. S. Mahesh}
	\email{mahesh.ts@iiserpune.ac.in}
	\author{C. S. Sudheer Kumar}
	\email{sudheer.kumar@students.iiserpune.ac.in}
	\author{Udaysinh T. Bhosale}
	\email{uday.bhosale@iiserpune.ac.in}
	\affiliation{Department of Physics and NMR Research Center,\\
		Indian Institute of Science Education and Research, Pune 411008, India}

	\begin{abstract}
		{
In conventional NMR experiments, the Zeeman energy gaps of the nuclear spin ensembles are much lower than their thermal energies, and accordingly exhibit tiny polarizations. Generally such low-purity quantum states are devoid of quantum entanglement.  However, there exist certain nonclassical correlations which can be observed even in such systems.  In this chapter, we discuss three such quantum correlations, namely, quantum contextuality, Leggett-Garg temporal correlations, and quantum discord.  In each case, we provide a brief theoretical background and then describe some results from NMR experiments.
		}
	\end{abstract}
	
	\maketitle

\begin{center}
\textit{`Correlations cry out for explanation'} - J. S. Bell in \textit{Speakable and Unspeakable in Quantum Mechanics}, Cambridge university press (1989).
\end{center}

\section{Introduction}
 Quantum physics is known for many nonintuitive phenomena including certain classically forbidden correlations.
 To study and understand these mysterious quantum correlations we require a suitable testbed.  Nuclear Magnetic Resonance (NMR) \cite{NMR_Abragam_book,Levitt_Spindynabook}
 of an ensemble of molecular nuclei in bulk liquids/solids form a convenient testbed even at room temperatures \cite{Corynmr_1st}.  
 The weakly perturbed nuclear spins in such systems can store quantum superpositions for long durations ranging from seconds to minutes.  In addition, excellent unitary controls via radio-frequency pulses allow precise manipulations of spin-dynamics.
 Even though one can not have local addressability of individual spins, and one works with the spin-ensemble as a whole, it is still possible to study many of the quantum correlations, namely contextuality, temporal correlation, discord etc.  The ensemble measurements are often sufficient, since many of the quantum correlations can be evaluated via expectation values.  However, at room temperature there is little entanglement in conventional NMR systems \cite{PhysRevLett.83.1054}.  In fact, this makes NMR a good candidate for studying quantum correlations without entanglement.  
 
 \begin{figure}[h]
 	\begin{center}
 		\includegraphics[trim = 5cm 5cm 2cm 2cm, clip, width=10cm ]{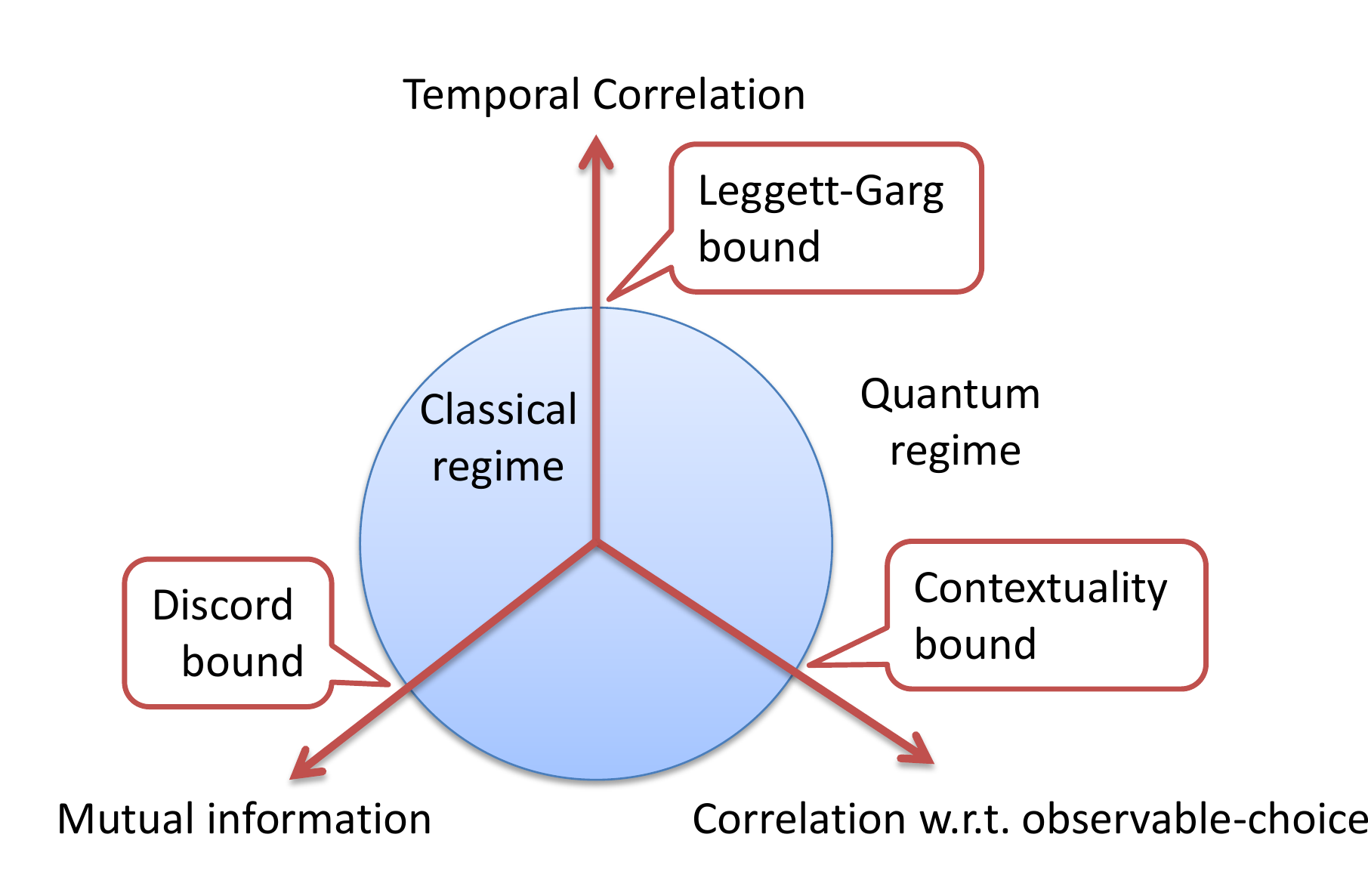}
 		\caption{Various correlations and corresponding bounds distinguishing quantum regime from classical.}
 		\label{overview}
 	\end{center}
 \end{figure}

In the following sections we are going to review some NMR experiments investigating quantum contextuality, Leggett-Garg inequality, and quantum discord. For the sake of completeness, we have provided a brief theoretical background in each case.

\section{Quantum Contextuality}
As the name suggests, outcome of a quantum measurement in general depends on the context i.e., measurement-setting, arrangement, situation, circumstance, etc. Quantum contextuality (QC) states that the outcome of a measurement depends not only on the system and the observable being measured, but also on the context of the measurement, i.e., on other compatible observables which are measured along with \cite{peres_context_1pg,quant_theory_peres,KS}. QC signifies a mysterious nonclassical correlation between  measurement outcomes corresponding to distinct observables.  
One consequence of QC is violation of Bell's inequality \cite{Cabello_state_ind_context,Bell_vio_lopholefree_1pt3km}, which
 has challenged the most cherished tenet of special theory of relativity, i.e., locality.

Peres explained quantum contextuality using a pair of electrons in a singlet state $(\ket{01}-\ket{10})/\sqrt{2}$ \cite{peres_context_1pg}. Suppose we measure a Pauli observable $\sigma_{i\alpha}$, where $\alpha \in \{x,y,z\}$, on the $i$th particle, and obtain an outcome $\alpha_i = \pm 1$. For the singlet state, the result of measuring $\sigma_{1x}\sigma_{2x}$ is $x_1x_2 = -1$ since $\expv{\sigma_{1x}\sigma_{2x}} = -1$. Similarly, $y_1y_2 = -1$ .  However,
if one measures $\sigma_{1x}\sigma_{2y}$ followed by 
 $\sigma_{1y}\sigma_{2x}$ one would obtain the outcome $x_1y_2y_1x_2 = -1$ since $\expv{\sigma_{1x}\sigma_{2y} \sigma_{1y}\sigma_{2x}} = \expv{\sigma_{1z}\sigma_{2z}} = -1$, which is in contradiction with $x_1x_2=y_1y_2=-1$.  
 
 Later Mermin \cite{Mermin} generalized quantum contextuality to a state-independent scenario. Consider a pair of spin-1/2 particles and a set of nine Pauli-observables arranged in the following fashion:
\begin{eqnarray}
 \begin{array}{|c|c|c||c|}
\hline
\sigma_{1z} & \sigma_{2z} & \sigma_{1z}\sigma_{2z} & +\mathbbm{1}\\
\hline
\sigma_{2x} & \sigma_{1x} & \sigma_{1x}\sigma_{2x}  & +\mathbbm{1} \\
\hline
\sigma_{1z}\sigma_{2x} & \sigma_{1x}\sigma_{2z} &  \sigma_{1y}\sigma_{2y} & +\mathbbm{1}\\
\hline
\hline
+\mathbbm{1} & +\mathbbm{1} & -\mathbbm{1} & \\
\hline
 \end{array}.
 \label{Merminsquare}
\end{eqnarray}
Here the last column (row) lists the product along the row (column).  
In this arrangement, all the operators along any row, or any column, mutually commute and therefore they can be measured sequentially or simultaneously without any mutual disturbance.
Whatever may be the state of the spin-pair, if one measures the three consecutive observables along any row one would obtain the outcome $+1$, the only eigenvalue of $\mathbbm{1}$.  Similarly, if one measures along  first or second column one would obtain  $+1$.  On the other hand, choosing observables along the last column will lead to an outcome $-1$.  However, no assignment of $\pm 1$ values to individual measurements of all the nine observables can satisfy the above joint-measurement outcomes, indicating that such noncontextual preassignments of measurement outcomes is incompatible with quantum physics.

\subsection{Contextuality studies using NMR systems} 
The first demonstration of contextuality in NMR systems was reported by Moussa \textit{et. al} \cite{Moussa}. Using a solid state NMR system, they evaluated the state independent inequality \cite{Cabello_state_ind_context}
\begin{eqnarray}
\beta = \expv{\pi_{r_1}}+\expv{\pi_{r_2}}+\expv{\pi_{r_3}}+
\expv{\pi_{c_1}}+\expv{\pi_{c_2}}-\expv{\pi_{c_3}} \le 4
\label{cabello}
\end{eqnarray}
where $\expv{\pi_{r_i}}$  are the expectation values obtained when all the observables along the $i$th row  of matrix in \ref{Merminsquare} are measured. Similarly
$\expv{\pi_{c_j}}$ is the expectation value for measurements along the $j$th column.
Exploiting the state independent property, they initialized the system in the maximally mixed state and obtained the value $\beta = 5.2\pm 0.1$.  While the result is in agreement with the quantum bound which is $\beta \le 6$, it strongly violates the inequality in \ref{cabello}. 

Later, Xi Kong \textit{ et. al} demonstrated QC by a single three level system in a NV center setup \cite{spin_1_QC}. More recently, Dogra \textit{ et. al}  demonstrated QC using a qutrit (spin-1) NMR system with a quadrupolar moment, oriented in a liquid crystalline environment. Using a set of 8 traceless observables (Gell-Mann matrices) and an inequality derived based on a noncontextual hidden variable (NCHV) model, they observed a clear violation of the NCHV inequality \cite{Contextuality_KDorai}.

\subsubsection*{Contextuality via psuedo spin mapping}
Su \textit{et. al.} \cite{Cont_theory} have theoretically studied QC of eigenstates of one dimensional quantum harmonic oscillator ($1$D-QHO) by introducing
two sets of pseudo-spin operators,
\begin{eqnarray}
{\bf \Gamma } = (\Gamma_x,\Gamma_y,\Gamma_z),~~
{\bf \Gamma '} = (\Gamma_x',\Gamma_y',\Gamma_z') \nonumber
\end{eqnarray}
with components,
\begin{eqnarray}
\Gamma_x = \sigma_x \otimes \mathbbm{1}_2,
\Gamma_y = \sigma_z \otimes \sigma_y,
\Gamma_z = -\sigma_y \otimes \sigma_y,  
\Gamma_x' = \sigma_x \otimes \sigma_z,
\Gamma_y' = \mathbbm{1}_2 \otimes \sigma_y,
\Gamma_z' = -\sigma_x \otimes \sigma_x,
\label{Gammas defined}
\end{eqnarray}
where $ \mathbbm{1}_2 $ is $2\times 2$ identity matrix. 
Defining the dichotomic unitary observables,
\begin{eqnarray}
A=\Gamma_x, ~~~ 
B=\Gamma_x' \cos \beta  + \Gamma_z' \sin \beta,  
~~~ 
C=\Gamma_z, 
~~~ 
D=\Gamma_x'\cos \eta  +  \Gamma_z'\sin\eta,
\label{A,B,C,D defined}
\end{eqnarray}
they setup  Bell-Clauser-Horne-Shimony-Holt (Bell-CHSH) inequality \cite{quant_info_neilson_chuang},
\begin{eqnarray}
\mathrm{\textbf{I}}= \expec{AB} + \expec{BC} + \expec{CD} - \expec{AD} \leq 2.
\label{BellCHSHineqlty}
\end{eqnarray}
However, the quantum bound was shown to be $\mathrm{\textbf{I}}_Q \le 2\sqrt{2}$, clearly violating the above
Bell-CHSH inequality and thus exhibiting QC of QHO.

Katiyar \textit{et. al.} carried out an NMR investigation of this inequality by mapping the QHO eigenstates to the spin-states of a 2-qubit system (with an additional ancilla qubit) \cite{QCQHOcssk}.
Using the Moussa protocol \cite{Moussa} (described in the next section) to extract the joint-expectation values in the inequality \ref{BellCHSHineqlty}, they obtained $\textbf{I}_Q \approx 2.4 \pm 0.1$.
Although decoherence limited the experimental value to below the quantum bound ($\textbf{I}_Q \le 2.82$),  it is clearly above the classical bound ($\textbf{I} \le 2$) and therefore establishes QC of 1D-QHO.

Thus, we observe that even when a system is in a separable state, measuring nonlocal observables leads to violation of Bell-CHSH inequality \cite{Cabello_NCHV_ineqlty}. 

\section{Temporal Correlations}
Bell's inequalities (BI) are concerned with how two systems (each with a dimension of at least 2) are correlated over space, where as the Leggett-Garg inequality (LGI) is concerned with the correlation of a single system (with a dimension of at least 2), with itself at different time instants. While the former deals with context of the measurement, the latter deals with a temporal context. 

LGI is based on the following two assumptions: 
\begin{itemize}
\item[1.] Macroscopic realism (MR): A macroscopic system, with two or more macroscopically distinct states available to it, exists in one of these states at any given point of time.
\item[2.]  Noninvasive measurability (NM): It is possible to determine the state of the system with arbitrarily small perturbation to its future dynamics \cite{LG,LGI_review}.
\end{itemize}

Although the original motivation of Leggett and Garg was to test the existence of quantumness even at a macroscopic level, most of the violations of LGI reported so far are on microscopic systems  \cite{LGI_review}.  The LGI violations in such systems were either due to invasive measurement or the system being in microscopic superpositions.  Even though LGI violation in a macroscopic system such as a superconducting qubit has been reported \cite{LeggettGargPalacios2010}, the existence of macroscopically distinct states in such a system is not clear \cite{LGI_review}. 
Other experimental works on LGI include Nitrogen-Vacancy centers \cite{Waldherr2011,George2013},  photonic systems \cite{Dressel2011}, electron interferometers \cite{LeggettGargEmaryClive},  superconducting qubit
 \cite{LeggettGargGroen2013}, and more recently in neutrino oscillations \cite{LeggettGargFormaggio}.
  Recent theoretical extensions of LGI include its entropic formulation\cite{ELGI_UshaDevi}
 and LGI in a large ensemble of qubits \cite{LeggettGargLambertNeill}. 
   The violation of the former was recently observed using NMR experiments by Katiyar {\it et. al.}\cite{ELGI_TSM}.
 LGI is also studied for a system of qubits 
 coupled to  a thermal environment \cite{LeggettGargLobejkoMarcin}. For more details reader can refer to the review
 \cite{LGI_review}. LGI violation in a 3-level NMR system has also been reported recently
  \cite{LeggettGargHemant2016}.
 
 In the following we provide a brief theoretical as well as experimental review of LGI in the context of NMR. 

\begin{figure}[h]
	\begin{center}
		\includegraphics[trim = 1.5cm 5cm 0cm 0cm, clip, width=17cm ]{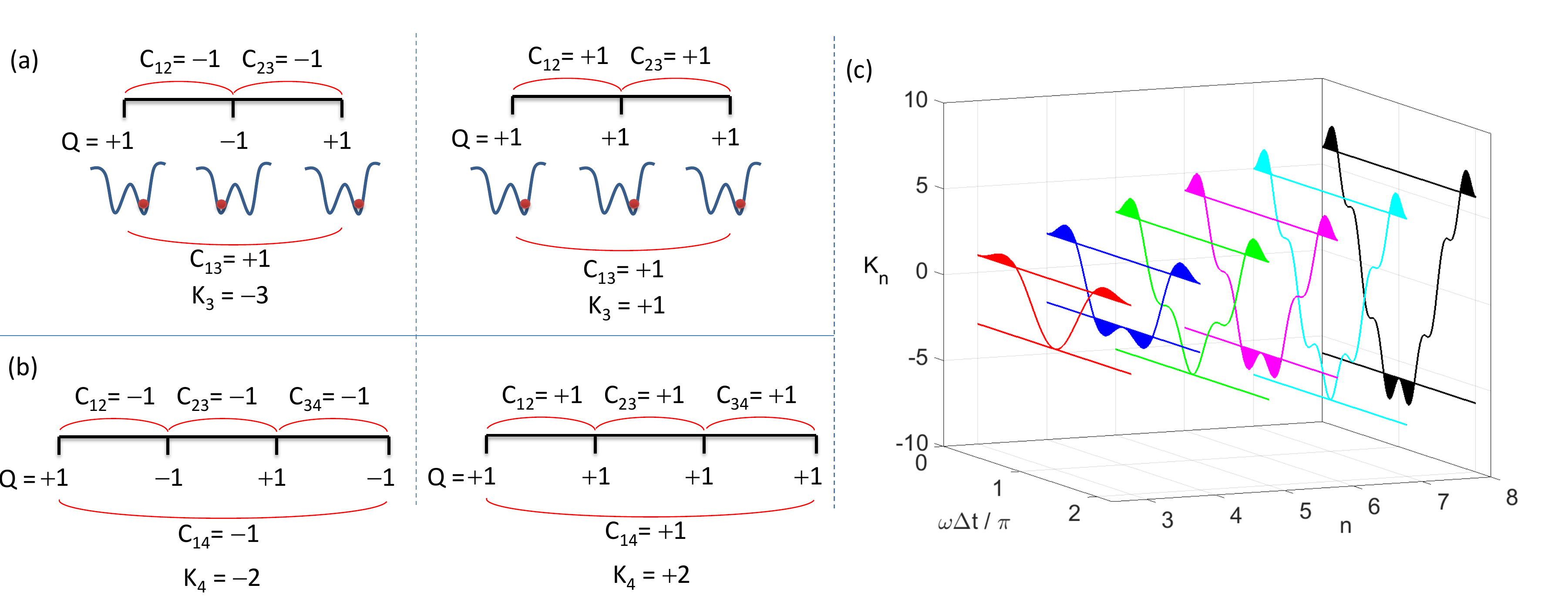}
		\caption{Extreme values of TTCCs for a classical particle in a double-well potential for the cases of (a) three-time measurement and (b) four-time measurement.  The left and right columns illustrate minimum and maximum values of $K_n$-strings respectively. (c) $K_n$ versus $n$ and $\omega \Delta t/\pi$ for a single qubit. The filled regions indicate LGI violations.}
		\label{kstring}
	\end{center}
\end{figure}

\subsection{Leggett-Garg string}
Consider a system (the `target') evolving under some Hamiltonian.  Let $\mathbbm{Q}$  be a dichotomic  observable  with eigenvalues $Q=\pm 1$, and let $Q(t_i)$ denotes the measurement outcome at time $t_i$.
Repeating these measurements a large number of times we obtain the two-time correlation coefficient (TTCC)  $C_{ij}$ for each pair: 
\begin{eqnarray}
C_{ij} = \lim\limits_{N\rightarrow\infty}\frac{1}{N} \sum _{r=1}^{N}Q_{r} \left( t_i \right)\cdot Q_r \left(t_j \right)=\expec{Q\left( t_i \right)\cdot Q\left(t_j \right)},
\end{eqnarray}
where $r$ is the trial number. Finally, the values of these coefficients are to be substituted in the \textit {n}-measurement LG string given by:
\begin{eqnarray}
K_{n} = C_{12}+ C_{23} + C_{34} +....+ C_{(n-1)n} - C_{1n}.
\end{eqnarray}
Each TTCC $C_{ij}$ is bounded by a maximum value of $+1$, corresponding to a perfect correlation, and a minimum value of $-1$, corresponding to a perfect anti-correlation.  $C_{ij}=0$ indicates no correlation. Thus, the upper bound for $K_{n}$ consistent with \textit{macrorealism} comes out to be $(n-2)$, while the lower bound is $-n$ for $odd$ $n$, and $-(n-2)$ for $even$ $n$ (see Fig. \ref{kstring}(a) and (b)). With these considerations  LGI reads $-n \le K_n \le (n-2) \;\; \mathrm{for \; odd} \;n \mathrm{,\; and} 
-(n-2) \le K_n \le (n-2) \;\; \mathrm{for \; even} \; n$.

In the following, we consider the case of a single qubit, namely a spin-1/2 nucleus precessing in an external static magnetic field.

\subsection{Violation of LGI with a single qubit}
A spin-1/2 nucleus precessing in an external magnetic field along $z$-axis has the following Hamiltonian: $\frac{1}{2}\omega \sigma_z$, where $\omega$ is the Larmor frequency. Let $\sigma_x$ be the dichotomic observable \cite{LGITSM}. 
Starting from the definition of TTCCs,
we obtain for an arbitrary initial state $\rho_0$ \cite{Brukner_multilevelLGI_Luders,tempCHSH_bound_fritz}, 
\begin{eqnarray}
C_{ij}=\left 
\langle \sigma_x\left(t_i\right) \sigma_x\left(t_j\right) \right\rangle =\cos \left\{ \omega (t_j-t_i) \right\}.
\label{Cij_theory}
\end{eqnarray} 
Dividing the total duration
from $t_{1}$ to $t_{n}$ into $(n - 1)$ parts each of length $\Delta t$, we can express the LG string consistent with equation \ref{Cij_theory} as
\begin{eqnarray}
K_{n}=(n-1)\cos\{\omega\Delta t\} - \cos\{(n-1)\omega\Delta t\}.
\label{kn}
\end{eqnarray}

Fig. \ref{kstring}(c) illustrates $K_n$ curves for $n=3$ to $8$ and for $\omega \Delta t \in [0,2\pi]$.  The classical bounds in each case are shown by horizontal lines.  As indicated by the filled areas, LGI is violated for each value of $n$ at specific regions of $\omega \Delta t$.  Quantum bounds of $K_3$ are $-3$ and $+1.5$ and that for $K_4$ are $-2\sqrt{2}$ and $+2\sqrt{2}$, and so on.
In the following we discuss an experimental protocol for evaluating the LG strings.

\subsection{Moussa protocol}
\begin{figure}[h]
	\begin{center}
		\includegraphics[trim = 0.5cm 1cm 0.5cm 0.5cm, clip, width=12cm ]{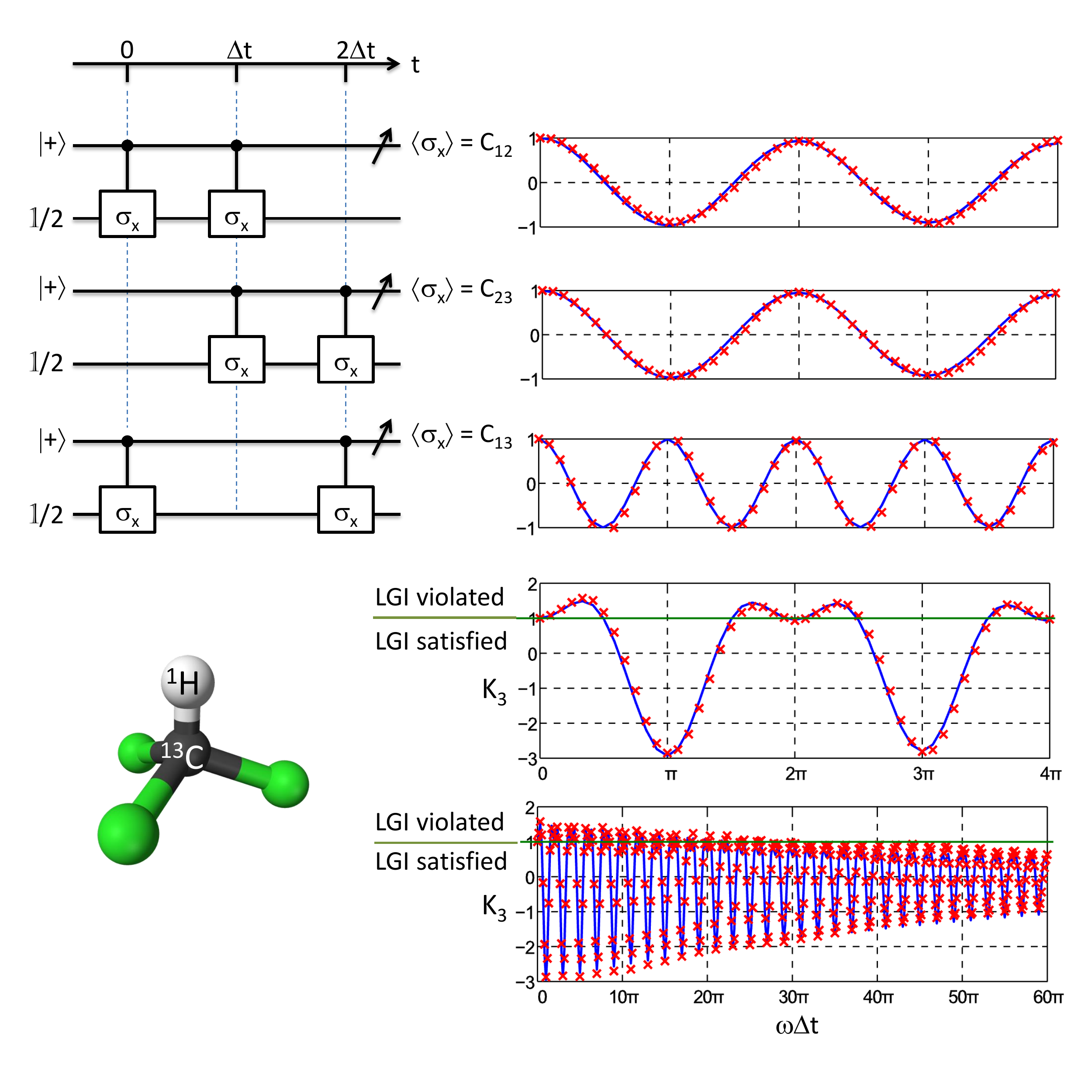}
		\caption{Moussa circuits (left) to extract TTCCs for the three-measurement case and the experimental results (crosses in the right) of $C_{ij}$ and $K_3$ obtained with $^1$H (ancilla) and $^{13}$C (system) spins of chloroform (molecular structure shown in bottom-left).  Both short-time and long-time behavior of $K_3$ are shown.  Here smooth curves are drawn with with theoretical expression (Eq. \ref{Cij_theory}) along with an appropriate decay factor.  
		Parts of this figure are adapted from \cite{LGITSM}.
}
		\label{lginmr}
	\end{center}
\end{figure}

As described before, one needs to extract TTCCs in a way as noninvasive as possible.  One way to achieve this is by using  
an ancilla qubit and employing Moussa protocol (Fig. \ref{lginmr}).  It involves preparing the ancilla in $\ket{+}$ state (an eigenstate of $\sigma_x$; or a pseudopure state $(1-\epsilon)\mathbbm{1}/2+\epsilon \proj{+}$) followed by a pair of CNOT gates separated by the delay $t_j-t_i$.  Finally $\sigma_x$ observable of the ancilla qubit is measured in the form of transverse magnetization which reveals the corresponding TTCC \cite{Moussa}:
\begin{equation}
\expv{\sigma_x}_\mathrm{ancilla} = \tr[\rho_s \sigma_x(t_i)\sigma_x(t_j)] = C_{ij},
\end{equation}
where $\rho_s = \mathbbm{1}/2$ is the initial state of the system qubit.

The Moussa circuits are easy to implement using a two-qubit NMR system \cite{LGITSM,Souza_LGI}.  Athalye \textit{et. al.} \cite{LGITSM} have used $^{13}$C and $^1$H spins of $^{13}$C-Chloroform as system and ancilla qubits respectively and found a clear violation of LGI by more than 10 standard deviations at short time scales.  However, with longer time scales, the TTCCs decayed resulting in a gradual reduction in the violation, and ultimately satisfying the LGI bounds.

More recently, Knee \textit{et. al.} \cite{LGI_knee_noninvasive} have used ideal negative result measurements (INRM) to extract TTCCs  noninvasively.  The method involves two sets of experiments - one with CNOT and the other with anti-CNOT.  In the former, the system qubit is unaltered if the ancilla (control-qubit) is in state $\ket{0}$, while in the latter, the system is unaltered if the ancilla is in state $\ket{1}$.  Postselecting the subspaces wherein the system is unaltered is considered to be more noninvasive \cite{LGI_knee_noninvasive}.  Using nuclear and electronic spins (in an ensemble of phosphorous donars in silicon) as system and ancilla, Knee \textit{et. al.}  demonstrated LGI violation with INRM \cite{LGI_knee_noninvasive}.

\subsection{Entropic Leggett-Garg inequality (ELGI)}
In 2013, Usha Devi {\it et. al.} \cite{ELGI_UshaDevi} have formulated the entropic Leggett-Garg inequality in which they place bounds on amount of information associated with a noninvasive measurement of a macroscopic system.  The amount of information stored in a classical observable $\mathbbm{Q}(t_i)$ at time $t_i$ is given by the Shannon entropy,
\begin{eqnarray}
H(\mathbbm{Q}(t_i)) = -\sum_{Q(t_i)} P(Q(t_i)) \log_2 P(Q(t_i)),
\end{eqnarray}
where $P(Q(t_i))$ is the probability of the measurement outcome $Q(t_i)$ at time $t_i$.
The conditional entropy $H(\mathbbm{Q}(t_j) \vert \mathbbm{Q}(t_i))$ is related to the joint-entropy 
\begin{eqnarray}
H(\mathbbm{Q}(t_j),\mathbbm{Q}(t_i)) = -\sum_{Q(t_i),Q(t_j)} P(Q(t_i),Q(t_j)) \log_2 P(Q(t_i),Q(t_j))
\end{eqnarray}
by Bayes' theorem, i.e.,
\begin{eqnarray}
H(\mathbbm{Q}(t_j) \vert \mathbbm{Q}(t_i)) = H(\mathbbm{Q}(t_i),\mathbbm{Q}(t_j))-H(\mathbbm{Q}(t_i)).
\end{eqnarray}
For $n$ measurements performed at equal intervals $\Delta t$,
we denote $h(\Delta t) = H(\mathbbm{Q}(\Delta t) \vert \mathbbm{Q}(0)) = 
H(\mathbbm{Q}(2\Delta t) \vert \mathbbm{Q}(\Delta t)) = 
\cdots$, and $h((n-1) \Delta t) = H(\mathbbm{Q}((n-1) \Delta t) \vert \mathbbm{Q}(0))$.
By setting up a quantity called information deficit
\begin{eqnarray}
{\mathcal D}_n = \frac{(n-1)h(\Delta t) - h((n-1)\Delta t  )}{\log_2(2s+1)},
\end{eqnarray}
where $2s+1$ is the number of distinct states (where $s$ is spin number),
Usha Devi \textit{et. al.} proved that ${\mathcal D}_n \ge 0$ for classical systems.  

\begin{figure}
	\begin{center}
		\includegraphics[trim = 0cm 0cm 0cm 0cm, clip, width=9cm ]{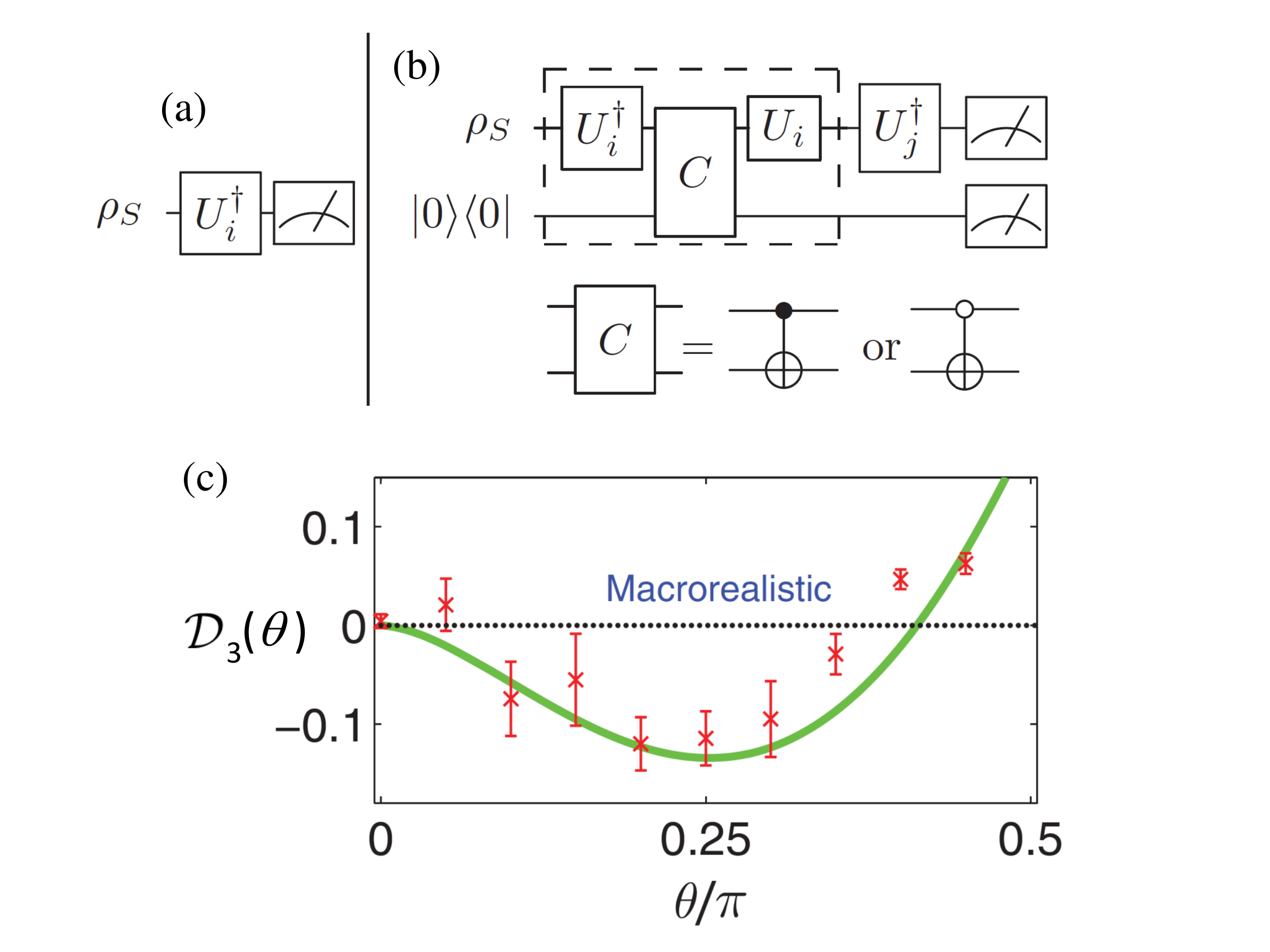}
		\caption{(a,b) The quantum circuits for extracting single-time and joint probabilities. Here $U_i^\dagger$ denotes the back-evolution of the system in the computational basis which is equivalent to having the dynamical observable $\mathbb{Q}(t_i)$. (c) Experimental information deficit (crosses with errorbars) compared to theoretical values (solid curve) for a spin-1/2 particle.  The dashed line indicates the macrorealistic bound.  Here $\theta = (n-1)\omega \Delta t$.
			Parts of this figure are adapted from \cite{ELGI_TSM}. 
			}
		\label{elgiresults}
	\end{center}
\end{figure}

The experimental violation of ELGI was 
first demonstrated by Katiyar \textit{et. al.} \cite{ELGI_TSM} again using $^{13}$C-Choroform as the two-qubit register.  The single-time probability $P(Q(t_i))$ and the joint probabilities $P(Q(t_i),Q(t_j))$ are extracted using the circuits shown in Fig. \ref{elgiresults}(a) and (b) respectively.  Note that an ancilla spin is used to extract  joint probabilities with the help of INRM procedure applied to the first measurement.
The results displayed in Fig. \ref{elgiresults}(c), indicate a clear violation of ELGI by four standard deviations.

\section{Quantum Discord}
In the early days of quantum information and quantum computation it was shown that
entanglement is the key resource to perform various tasks \cite{Horodeckirpm}.
However, it was later realized that quantum correlations beyond entanglement are also useful for 
quantum information processing \cite{Knill1998,Bennett1999,Niset2006,Horodecki2005}.
It was shown theoretically \cite{PhysRevLett.83.1054,DavidMeyer2000} as well as experimentally \cite{Lanyon2008} 
that some tasks can be made efficient even with separable 
states, but with non-zero quantum correlations.
Thus, quantifying the quantum correlation becomes important, and it can be achieved by using measures such as discord \cite{Ollivier,vedral} and geometric discord 
\cite{Dakic,Luo_geometric,PaoloGiorda2010}.
For more details on the topic of quantum correlations one may refer to the reviews in \cite{ModiRMP,ModiOpenSyst,Zhang2012,Horodecki2013,Streltsov,AdessoReview}.

Discord has also been studied in the ground state of certain spin chains particularly close to quantum phase 
transitions \cite{Sarandy2013}. Signatures of chaos in the dynamics of quantum discord are found using the 
model of the quantum kicked top \cite{VaibhavMadhok2015}. Quantum critical behavior in the anisotropy $XY$ spin 
chain is studied using geometric discord \cite{Cheng2012}.

It is believed that discord is a resource behind the efficiency of the DQC-1 model 
\cite{Knill1998,DavidMeyer2000,Passante2009,Maziero2009,Animesh}.  Quantum advantage with no entanglement but with non-zero quantum discord has been demonstrated in single-photon states \cite{Maldonado2016}.  Quantum discord has also been estimated in optical systems using mixed states \cite{Lanyon2008} and in an anti ferromagnetic Heisenberg compound \cite{Mitra2015}. 

Non-zero quantum discord in NMR systems has been observed by many researchers \cite{laflammeoct11,serraaug11,serradiscordquad,katiyarDiscord}.
For various theoretical and experimental aspects of quantum discord and related measures reader
can refer to review \cite{Celeri2011}. 
Investigations on the evolution of quantum discord under decoherence\cite{DiscordNMR_Oliveira} and under decoherence-suppression sequences \cite{katiyarDiscord} have also been reported.  
In the following we briefly describe some aspects related to discord and geometric discord.

\subsection{Discord and mutual information}
\begin{figure}
	\begin{center}
		\includegraphics[width=12cm]{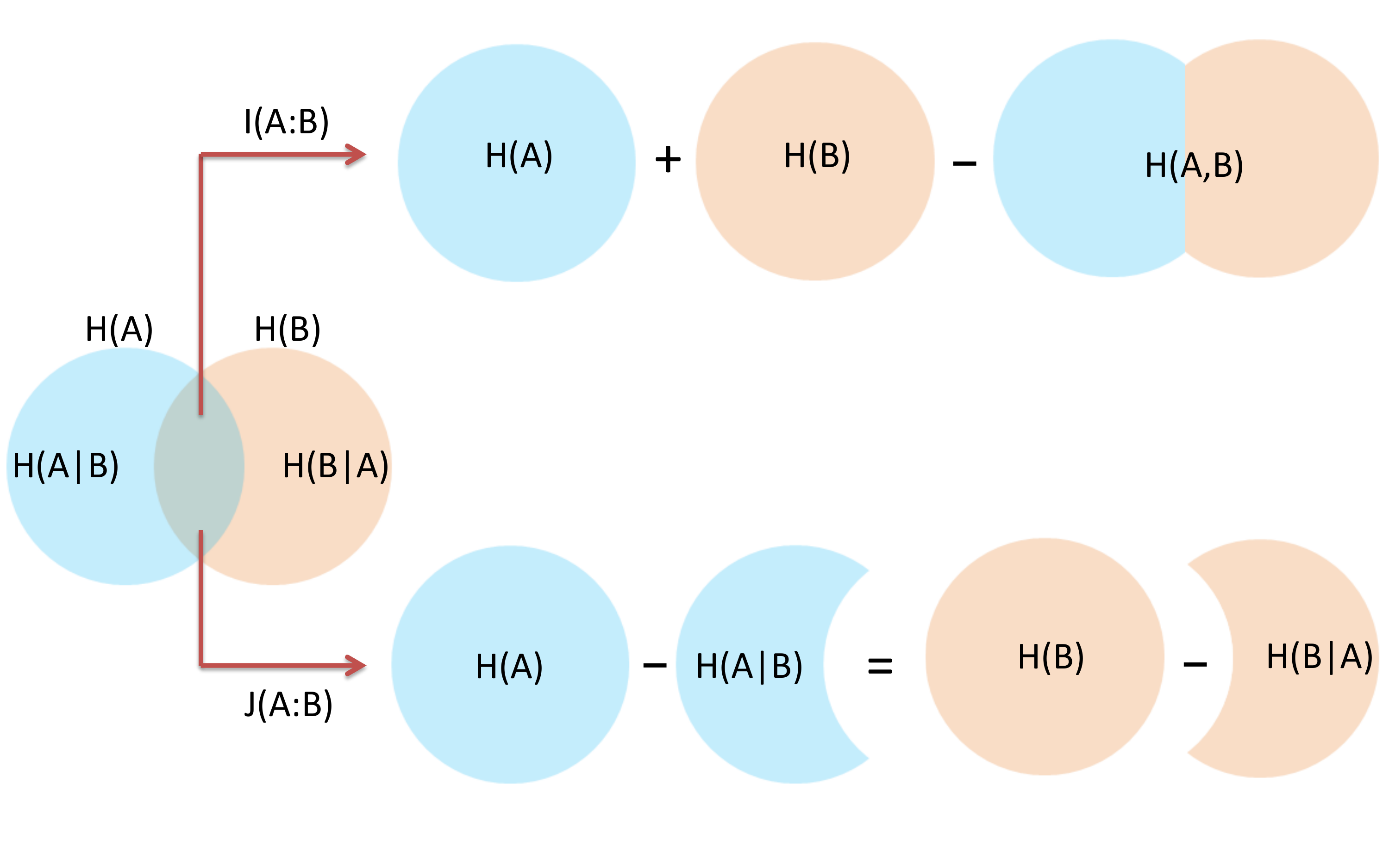}
		\caption{Venn diagram representing total information $H(A,B)$,
			individual informations $\left(H(A), H(B)\right)$, the conditional
			information $\left( H(A|B), H(B|A)\right)$, and the mutual information
			$I(A:B) = J(A:B)$ in classical information theory.}
		\label{mutual}
	\end{center}
\end{figure} 

Mutual Information $I(A:B)$   is defined as the amount of information that is common to both the subsystems $A$ and $B$ of a 
bipartite system, and is given in terms of Shannon entropy
\begin{equation}
I(A:B)= H(A) + H(B) - H(A,B).
\label{eq5}
\end{equation}

It can be seen that mutual information is symmetric, i.e., $I(A:B) = I(B:A)$. 
Another classically equivalent expression based on Bayes rule can be obtained from Eq.~(\ref{eq5})
as follows:
\begin{eqnarray}
J(A:B) = H(A) - H(A|B) 
= H(A) - \sum_{i}p_{i}^{b}H(A|b=i) ~.
\label{eq6}
\end{eqnarray}
These expressions can be intuitively understood using 
Fig.~\ref{mutual}. 

In the quantum information theory, the von Neumann entropy gives the information content of a density matrix
and is defined as 
\begin{equation}
{\mathcal H}(\rho) = -\sum_{x}\lambda_{x}\log_{2}\lambda_{x},
\label{eq7}
\end{equation}
where $\lambda_{x}$'s are the eigenvalues of the density matrix $\rho$. Although the two expressions of mutual 
information given in Eqs.~(\ref{eq5}) and (\ref{eq6}) are equivalent in classical information theory
this is not the case in quantum information theory. The reason for this difference is that
the expression for mutual information given by Eq.~(\ref{eq6}) involves
measurements and its value depends on the measurement outcomes. Measurements in quantum theory
depends on the basis used and it changes the final state of the system. Henderson
and Vedral \cite{vedral} have proved that the total classical correlation
can be obtained as the maximum value of 
\begin{eqnarray}
{\mathcal J}(A:B) = {\mathcal H}(B) - {\mathcal H}(B|A) 
= {\mathcal H}(B) - \sum_{i}p_{i}^{a}{\mathcal H}(B|a=i) ~,
\label{eq8}
\end{eqnarray}
where the maximization is performed over all possible orthonormal measurement
bases $\{\Pi_{i}^{a}\}$ for $A$.
The quantum mutual information ${\mathcal I}(A:B)$ is defined in a way analogous to that of the classical mutual information, i.e., 
\begin{eqnarray}
{\mathcal I}(A:B) = {\mathcal H}(A)+{\mathcal H}(B)-{\mathcal H}(A,B).
\label{qmutualinfo}
\end{eqnarray}
Therefore, the non-classical correlations can be quantified as the difference
\begin{equation}
D(B|A) = {\mathcal I}(A:B) - \max_{\{\Pi_{i}^{a}\}}{\mathcal J}(A:B).
\label{disc}
\end{equation}
Ollivier and Zurek had called this difference as `\textit{discord}' \cite{Ollivier}.
Zero-discord states or ``classical'' states are the ones in which the
maximal amount of information about a subsystem can be obtained without disturbing
its correlations with the rest of the system.

It should be noted that discord is not a symmetric function in general, i.e. $D(B|A)$ and $D(A|B)$
can differ. Datta \cite{animesh_nullity} has proved that a given state
$\rho_{AB}$ satisfies $D(B|A)=0$ if and only if there exists a complete
set of orthonormal measurement operators on $A$ such that
\begin{equation}
\rho_{AB} = \sum_{i}p_{i}^{a} \Pi_{i}^{a}\otimes\rho_{B|a=i}.
\label{classical}
\end{equation}
When the first part of a general bipartite system is measured, the resulting density matrix is of the form 
given by Eq.~(\ref{classical}). Since the final state after
measurements is a classical state, one can extract the 
classical correlations from it. Thus, for any quantum state and every orthonormal measurement basis, there exists 
a classically correlated state. Maximization of ${\mathcal J}(A:B)$ gives the maximum classical correlation that
can be extracted from the system, and the remaining extra correlation is the quantum correlation.

\subsection{Evaluation of Discord}
Given a density matrix $\rho_{AB}$, one can easily construct the reduced density
matrices $\rho_{A}$ and $\rho_{B}$ of the individual subsystems.
Then the total correlation
${\mathcal I}(A:B)$ can be found using the quantum mutual information Eq.(\ref{qmutualinfo}). Maximization of ${\mathcal J}(A:B)$
to evaluate discord is nontrivial. The brute force method is to
maximize ${\mathcal J}(A:B)$ over as many orthonormal measurement bases as possible,
taking into account all constraints and symmetries.
Strictly speaking, this method gives a lower bound on  ${\mathcal J}(A:B)$ since the maximization may not be perfect.

While a closed analytic formula for discord does not exist for a general quantum state, analytical results are available for certain
special classes of states  \cite{girolami}. 
For example, Chen \textit{et. al.} have described analytical evaluation of discord for two qubit $X$-states under 
specific circumstances \cite{chen,TingYu2007,ARPRau2009,ARPRau2010,Fanchini2010}. 
Luo has given an analytical formula for discord of the Bell-diagonal states which are
a subset of the $X$-states \cite{luo}, and 
are defined as the states which are diagonal in the Bell
basis
\begin{equation}
\vert \psi^{\pm}\rangle = \frac{1}{\sqrt{2}}
( \vert 01\rangle \pm \vert 10\rangle ) ~,~~
\vert \phi^{\pm}\rangle = \frac{1}{\sqrt{2}}
( \vert 00\rangle \pm \vert 11\rangle ).
\end{equation}
The generic structure of a Bell-diagonal state is
$\rho_{BD} = \lambda_1 \vert \psi^-\rangle\langle \psi^- \vert +
\lambda_2 \vert \phi^-\rangle\langle \phi^- \vert +
\lambda_3 \vert \phi^+\rangle\langle \phi^+ \vert +
\lambda_4 \vert \psi^+\rangle\langle \psi^+ \vert$.
This state is separable iff it’s spectrum lies in $[0,1/2]$ \cite{RyszardHorodecki1996}.

Using only local unitary operations (so that the correlations remain unaltered), all Bell-diagonal states can 
be transformed to the form given by
\begin{equation}
\rho_{BD} = \frac{1}{4} \Big( \mathbbm{1} 
+ \sum_{j=1}^{3}r_{j}\sigma_{j}\otimes\sigma_{j} \Big),
\label{rhobd}
\end{equation}
where the real numbers $r_{j}$ are constrained such that all eigenvalues of
$\rho_{BD}$ remain in $[0,1]$. The symmetric form of $\rho_{BD}$ also
implies that it has symmetric discord, i.e., $D_{BD}(B|A) = D_{BD}(A|B)$.
Thus, the analytical formula for discord in this case is, using Eq.~(\ref{disc}),
\begin{eqnarray}
{D}_{BD}(B|A) &=& 2 + \sum_{i=1}^{4}\lambda_{i}\log_2\lambda_{i}
- \left(\frac{1-r}{2}\right)\log_{2}(1-r) 
- \left(\frac{1+r}{2}\right)\log_{2}(1+r),
\label{dbd}
\end{eqnarray}
where $r=\max\{|r_{1}|,|r_{2}|,|r_{3}|\}$.

A special Bell-diagonal state, i.e., when $\lambda_1=(1+3\epsilon)/4$ and 
$\lambda_2=\lambda_3=\lambda_4=(1-\epsilon)/4$, is known as the Werner state
\begin{eqnarray}
\rho_{W}(\epsilon) = \frac{1-\epsilon}{4}\mathbbm{1}
+ \epsilon \vert\psi^- \rangle\langle \psi^- \vert.
\label{werner}
\end{eqnarray}
It has entanglement iff $1/3 \leq \epsilon \leq 1$.
In this case $r_j=-\epsilon$ for $j=1$, $2$, $3$ and $r=\epsilon$. The discord using Eq.~(\ref{dbd}) is then given by
\begin{eqnarray}
D_{W}(\epsilon) = \frac{1}{4}
\log_2 \frac{(1-\epsilon)(1+3\epsilon)}{(1+\epsilon)^2}+ \frac{\epsilon}{4}
\log_2 \frac{(1+3\epsilon)^3}{(1-\epsilon)(1+\epsilon)^2} 
= \frac{\epsilon^2}{\ln 2} + O(\epsilon^3).
\label{dwerner}
\end{eqnarray}
This expression is plotted in Fig.~\ref{discordVspurity}.

\subsection{Geometric Discord}
Geometric discord is a form of Discord that is relatively easier to compute \cite{Dakic,Luo_geometric}.  In the following, we discuss
the case of two-qubit geometric discord \cite{Dakic,Huang2016}.
For every quantum state there exist a set of postmeasurement classical states ($\Omega_0$), and the geometric discord is 
defined as the distance between the quantum state ($\rho$) and the nearest classical state ($\chi$),
\begin{equation}
D^{G}(B|A) = \min_{\chi \in \Omega_0}\|\rho-\chi\|^2,
\end{equation}
where
$\|\rho-\chi\|^2 = {\rm Tr}[(\rho-\chi)^2]$ is the Hilbert-Schmidt quadratic norm.
Obviously, $D^{G}(B|A)$ is invariant under local unitary transformations.
Explicit and tight lower bound on the geometric discord for an arbitrary
$A_{m \times m} \otimes B_{n \times n}$ state of a bipartite quantum
system is available \cite{Luo_geometric,Hassan}. 
Protocols to determine lower bounds on geometric discord without tomography have also been discovered recently \cite{Rana,Hassan}.

Following the formalism of Dakic  \textit{et. al.} \cite{Dakic} analytical expression for the geometric discord for 
two-qubit states was obtained in \cite{RyszardHorodecki1996}.
The two-qubit density matrix in the Bloch representation is 
\begin{equation}
\rho = \frac{1}{4} \Big( \mathbbm{1} \otimes \mathbbm{1}
+ \sum_{i=1}^{3} x_{i}\sigma_{i}\otimes\mathbbm{1}
+ \sum_{i=1}^{3}y_{i} \mathbbm{1}\otimes\sigma_{i}
+ \sum_{i,j=1}^{3} T_{ij}\sigma_{i}\otimes\sigma_{j} \Big),
\label{bloch}
\end{equation}
where $x_{i}$ and $y_{i}$ represent the Bloch vectors for the two qubits,
and $T_{ij}={\rm Tr}[(\rho(\sigma_{i}\otimes\sigma_{j}))]$ are the components
of the correlation matrix. The geometric discord for such a state is
\begin{equation}
D^{G}(B|A) = \frac{1}{4}\left(\|x\|^2 + \|T\|^2 - \eta_{\rm max} \right),
\label{dg}
\end{equation}
where $\|T\|^2 = {\rm Tr}[T^\dagger T]$,\ and $\eta_{\rm max}$ is the largest eigenvalue of the matrix 
$\vec{x}\vec{x}^\dagger + TT^\dagger$. Explicit form of $\eta_{\rm max}$  and a remarkable tight lower bound on geometric discord are given in \cite{RyszardHorodecki1996}.

Using the transformed form of Bell-diagonal states as given in Eq.~(\ref{rhobd}) it can be seen that 
$x_i = y_i = 0$ and $T$ is a diagonal matrix with elements $T_{ii}=r_i$. Then the geometric discord is given as 
\begin{equation}
D^{G}_{BD}= \frac{1}{4}\left(\sum_{i=1}^{3} r_i^2 - \mbox{max}(r_1^2,r_2^2,r_3^2) \right).
\end{equation}

\begin{figure}
	\begin{center}
		\hspace*{-0.5cm}
		\includegraphics[angle=0,width=9cm,trim = 4cm 3cm 5cm 3cm,clip=,]{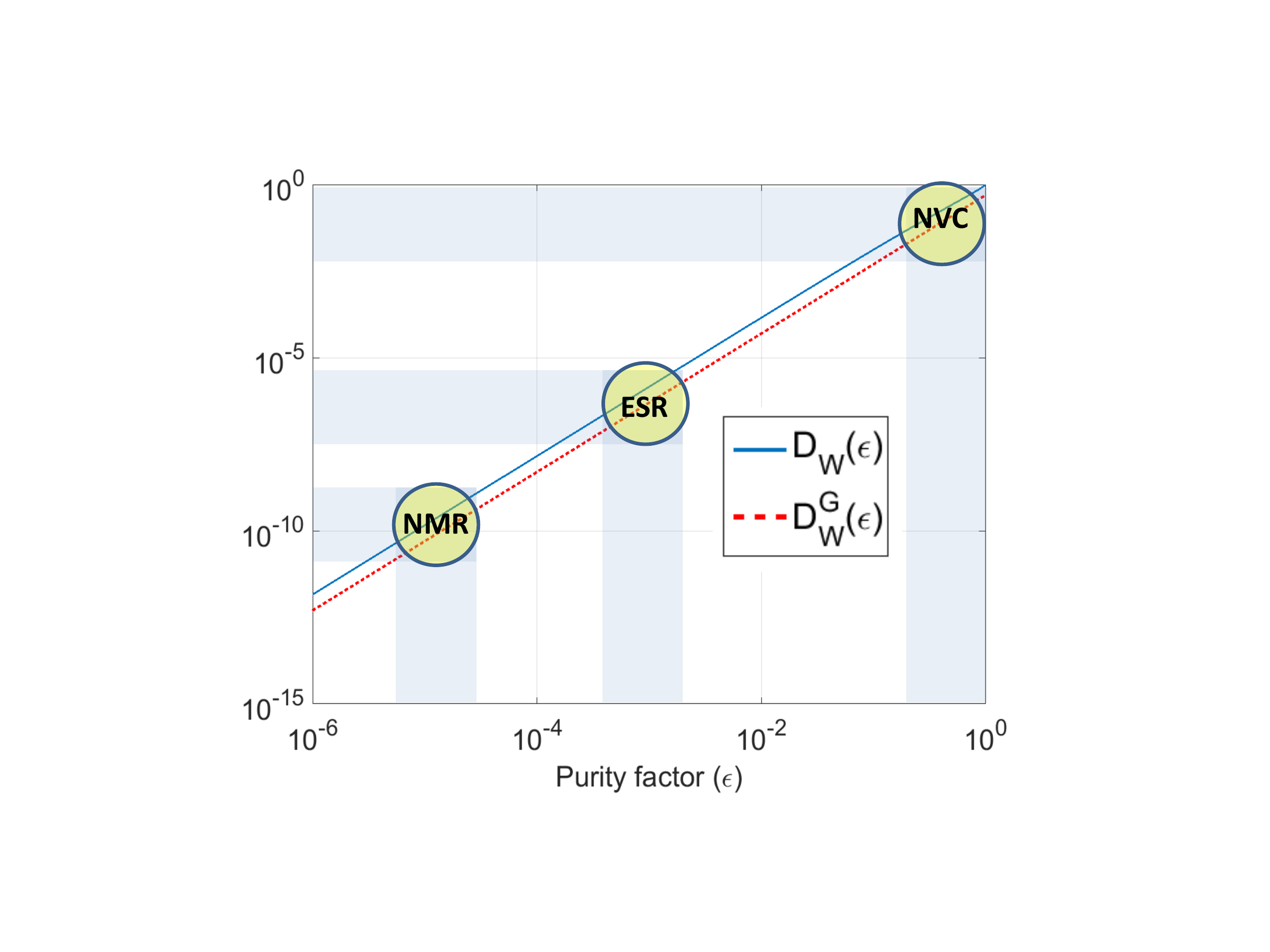}
		\caption{Discord ($D_W$) and geometric discord ($D_W^G$) of Werner state as a function of its purity factor $\epsilon$.  Typical ranges of purity and discord values for some spin-based architectures such as NMR, low-field ESR, and optically polarized electronic spin of nitrogen-vacancy center (NVC) are indicated.}
		\label{discordVspurity}
	\end{center}
\end{figure}

For the Werner state $r_i=-\epsilon$.
Then $\|T\|^2 = 3\epsilon^2$ and
all eigenvalues of $TT^\dagger$ are $\epsilon^2$, yielding
\begin{equation}
D^{G}_{W}(\epsilon) = \frac{1}{4}\left(3\epsilon^2-\epsilon^2 \right) = \frac{\epsilon^2}{2}.
\end{equation}
This expression is plotted versus the purity $\epsilon$ in
Fig.~\ref{discordVspurity}. Comparison with Eq.~(\ref{dwerner}) reveals that
discord and geometric discord are proportional for low-purity Werner
states. Also, the numerical difference between $D_{W}(\epsilon)$ and
$2 D^G_{W}(\epsilon)$ does not exceed 0.027 for all $\epsilon\in[0,1]$.
An analytical formula for symmetric geometric discord for two-qubit systems is given in
\cite{Shi2011} and geometric discord for qubit--qudit systems is given in \cite{Sai2012}.

\subsection{NMR studies of quantum discord}
\begin{figure}
	\begin{center}
		\hspace*{-0.5cm}
		\includegraphics[angle=0,width=9cm,trim = 7cm 4cm 5.5cm 2cm,clip=,]{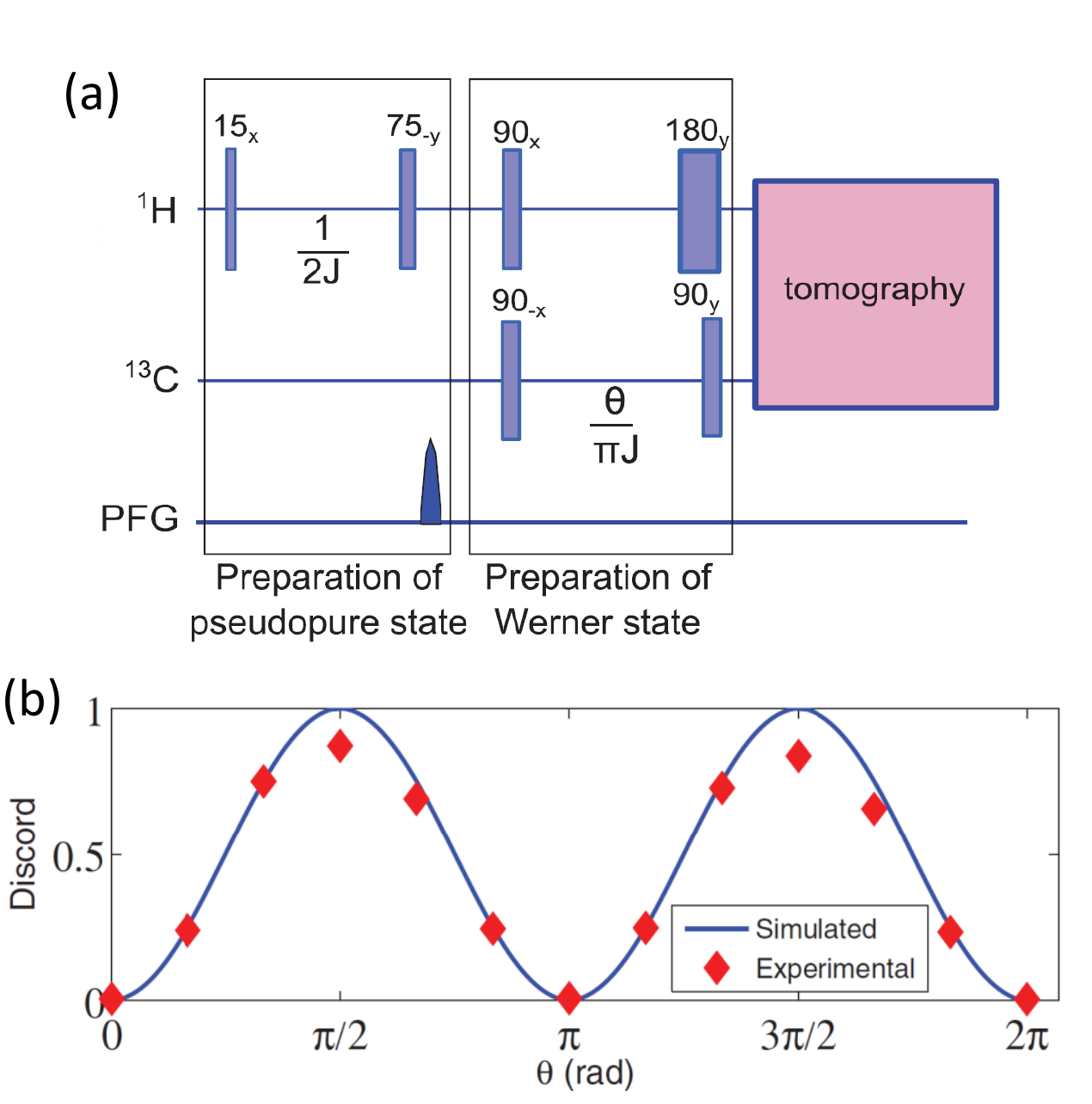}
		\caption{(a) NMR Pulse-sequence used by Katiyar \textit{et. al.} to prepare Werner state and measure discord and (b) experimental and simulated discord as a function of the nonlocal rotation $\theta$. Parts of this figure are adapted from \cite{katiyarDiscord}.}
		\label{DwCHCl3}
	\end{center}
\end{figure}

Katiyar \textit{et. al.} \cite{katiyarDiscord} have studied discord and its evolution in certain NMR systems.  After preparing the pseudopure state $\rho_0 = (1-\epsilon)\mathbbm{1}/2+\epsilon \proj{00}$ they  applied the pulse sequence shown in Fig. \ref{DwCHCl3}(a).  The initial  state $\rho_0$ is transformed into a Werner state when $\theta$ is set to an odd integral multiple of $\pi/2$.  Katiyar \textit{et. al.} measured quantum discord using extensive measurement method described earlier.  Fig. \ref{DwCHCl3}(b) displays discord as a function of $\theta$.  One can notice that discord is zero for the initial state $\rho_0$, grows with $\theta$ and reaches a maximum value at the Werner state.  This experiment demonstrates the existence of small, but non-zero,  nonclassical correlations in NMR systems even at room temperatures.

Maziero \textit{et. al.} studied the behavior of quantum discord under decoherence using an NMR testbed \cite{DiscordNMR_Oliveira}. They observed a sudden change in the behavior of classical and quantum correlations at a particular instant of time and found distinct time intervals where classical and quantum correlations are robust against decoherence.
Yurishchev \cite{DiscordNMR_Yurishchev2014} has analytically and numerically studied NMR dynamics of quantum discord in gas
molecules (with spin) confined in a closed nanopore. Kuznetsova and Zenchuk \cite{DiscordNMR_dimers_Kuznetsova20121029} have theoretically studied quantum discord in a pair of spin-1/2 particles (dimer) governed by the standard multiple quantum NMR Hamiltonian and shown the relation between discord and the intensity of the
second-order multiple quantum coherence in NMR systems. 

\section{Summary}
In this chapter, we have briefly discussed three types of quantum correlations, namely quantum contextuality, Leggett-Garg temporal correlations, and quantum discord.  In each case, we have surveyed a few NMR experiments. 

Exploiting the state independent nature of quantum contextuality, Moussa \textit{et. al.} \cite{Moussa} demonstrated that even a content-less maximally mixed-state ($\mathbbm{1}/4$) violates certain noncontextual hidden variable inequalities when subjected to quantum measurements of certain observables. 
Similarly, the violation of Leggett-Garg inequalities can be observed even in a two-level quantum system (while quantum contextuality is exhibited by a quantum system with at least three levels).
Hence, as demonstrated by Athalye \textit{et. al.} \cite{LGITSM} the violation of LGI is observable even in a spin-1/2 NMR system at room temperature.  Moreover,  Oliveira \textit{et. al.} \cite{DiscordNMR_Oliveira} and Katiyar \textit{et. al.} \cite{ELGI_TSM} showed the existence of nonzero discord in NMR systems.  

NMR has wide-ranging applications from spectroscopy to imaging, and quantum information testbed is the latest of them. Although NMR offers excellent control operations and long coherence times, highly mixed nature of spin-ensembles at room temperatures allows only separable quantum states.  In the absence of entanglement, does it have any resource for quantum information studies?  This question was answered in terms of above nonclassical correlations.  

\section*{Acknowledgments}
TSM acknowledges support from DST/SJF/PSA-03/2012-13 and CSIR 03(1345)/16/EMR-II. UTB acknowledges support from DST-SERB-NPDF (File Number PDF/2015/000506).

\bibliography{reference22013}

\end{document}